\documentclass[11pt,preprint]{aastex}
\usepackage{epsfig}

\def\etal{\it et al. \rm }
\begin{document}

\title{ARCHANGEL Galaxy Photometry System}

\author{James Schombert}
\affil{Department of Physics, University of Oregon, Eugene, OR 97403;
js@abyss.uoregon.edu}

\begin{abstract}

\noindent Photometry of galaxies has typically focused on small, faint systems due to
their interest for cosmological studies.  Large angular size galaxies, on
the other hand, offer a more detailed view into the properties of galaxies,
but bring a series of computational and technical difficulties that inhibit
the general astronomer from extracting all the information found in a
detailed galaxy image.  To this end, a new galaxy photometry system has
been developed (mostly building on tools and techniques that have existed
in the community for decades) that combines ease of usage with a mixture of
pre-built scripts.  The audience for this system is a new user (graduate
student or non-optical astronomer) with a fast, built-in learning curve to
offer any astronomer, with imaging data, a suite of tools to quickly extract
meaningful parameters from decent data.  The tools are available either by
a client/server web site or by tarball for personal installation.  The
tools also provide simple scripts to interface with various on-line
datasets (e.g. 2MASS, Sloan, DSS) for data mining capability of imaged
data.

\noindent As a proof of concept, we preform a re-analysis of the 2MASS Large Galaxy
Atlas to demonstrate the differences in an automated pipeline, with its
emphasis on speed, versus this package with an emphasis on accuracy.  This
comparison finds the structural parameters extracted from the 2MASS
pipeline is seriously flawed with scale lengths that are too small by 50\%
and central surface brightness that are, on average, 1 to 0.5 mags too
bright.  A cautionary tale on how to reduce information-rich data such as
surface brightness profiles.  This document and software can be found at
http://abyss.uoregon.edu/$\sim$js/archangel.

\end{abstract}

\section{Introduction}

The photometric analysis of large galaxies is a double edged sword.  While
increased resolution, compared to distant galaxies, provides avenues for
more detailed analysis of galaxy properties (to name a few: examination of
star formation regions, examination of core parameters to search for
massive blackholes, spiral arm analysis, isophotal irregularities that may
signal rings, bars or other secular evolution processes), it is a data
reduction fact that the larger number of pixels complicates the extraction
of simple parameters, such as total magnitude, mean surface brightness and
isophotal radius.  For example, the fact that the galaxy is spread over a
larger area of the sky means that the outer pixels have more sky luminosity
than galaxy luminosity, increasing the error in any galaxy value.  In
addition, increased angular size has frequently prevented a fair comparison
between distant and nearby galaxy samples simply because the techniques
used to extract parameters from nearby galaxies differ from those used on
small galaxies.

In general, the analysis of large galaxies (i.e. ones with many pixels)
requires a full surface photometry study of the isophotes and their shape.
For most extragalactic systems, the shape of choice is an ellipse.  The
astrophysics behind this assumption is that galaxy light traces stellar
mass, and stellar mass follows elliptical orbits as given by Kepler's 1st
law.  Certainly, this is true the case of early-type galaxies (elliptical
and S0's) as demonstrated by studies that examined the residuals from
elliptical isophotes (Jedrzejewski 1987).  This is also mostly true for
disk galaxies, although the lumpiness of their luminosity distribution due
to recent star formation increases the noise around each ellipse (see the
discussion in \S 2.4).  For dwarf irregular systems, any regular contours are
poor describers of their shape, thus an ellipse is used because it
is the simplest shape (aside from a circle) to describe an irregular
isophote.

Therefore, the analysis of large galaxies begins with the reduction of
their 2D images into 1D surface photometry as described by elliptical
isophotes.  In turn, the 1D profiles can be fit to various functions in
order to extract characteristic luminosities (stellar mass), scale lengths
and standard surface brightnesses (luminosity density).  When combined with
kinematic information, these three parameters form the Fundamental Plane
for galaxies, a key relationship for understanding the formation and
evolution of galaxies.

Aside from direct relevance to the Fundamental Plane, the need for better
galaxy photometric tools has also increased with the influx of quality HST
imaging.  Before HST, distant galaxies were mere point sources, but now
with WFPC2, ACS and NICMOS data, there is the need to perform full surface
photometric studies on a much larger volume of the Universe.  The sizes of
our database on the photometric structure of galaxies has increased a
thousandfold in the last 10 years, but most of the tools used to reduce
this new data are up to 20 years out-of-date.  Thus, the analysis of high
resolution space imaging data is far behind spectroscopic and high energy
data, not due to lack of interest, but due to the inadequacy of our 2D
analysis tools.

The goal of this software project (called the ARCHANGEL project for obtuse
historical reasons) has been to produce a series of proto-NVO type tools,
related to surface photometry, and develop a computing environment which
will extend the capability of individual observers (or entire data centers)
to perform virtual observatory science.  There is no attempt herein to
replace existing analysis packages (i.e. PyIRAF), but rather our goal is to
supplement existing tools and provide new avenues for data reduction as it
relates to galaxy photometry.  We hope that the fundamental components in
this package will provide the community with new methods to which they can
add their own ideas and techniques as well as provide learning environment
for new researchers.  In addition, there is growing amount of data by
non-optical astronomers as new space missions provide imaging data in
wavelength regions previously unexplored.  Thus, there is a new and growing
community of non-optical astronomers with 2D analysis needs that we hope to
serve.

\section{Package Philosophy}

The tools described herein are not intended to be a complete data reduction
package per say, but rather a set of basic modules that allows the user to
1) learn the procedures of galaxy photometry, 2) tailor the tools to their
particular needs, 3) begin an advanced learning curve of combining basic
modules to produce new and more sophisticated tools.  Turning raw data
(level 1 or 2 data) from the telescope (space or ground) into calibrated,
flattened images is the job of other, more powerful packages such as PyRAF.
The tools presented herein bridge the intermediate step between calibrated
data and astrophysically meaningful values.  Specifically, we are concerned
with the analysis of 2D images into 1D values, such as a surface brightness
profile, and further tabulation into final values, such as total luminosity
or scale length.

With respect to galaxy images, the numbers most often valued are
luminosity, scale length and luminosity density.  Unfortunately, due to the
extended nature of galaxies, the quality and accuracy of these values can
varying depending on the type of value desired.  For example, luminosities
can be extracted in metric form (luminosity within 16 kpc) or isophotal
(luminosity inside the 26.5 mag arcsecs$^{-2}$ isophote or the total
luminosity, an extrapolation to infinite radius.  Scale length can be
expressed as the radius of half-light or a formula fit to the luminosity
distribution (e.g. Seric function).  Luminosity density can be described
through a detailed surface photometry profile, or integrated as a mean
surface brightness within an isophote, or again a fitted curve such as an
exponential disk.  The tools provided by this project allow an
inexperienced user the capability to explore their dataset and extract
meaningful results while outlining the limitations to that data.

For the experienced researcher, these tools enhance their previous
background in data reduction and provide new, and hopefully, faster avenues
of analysis.  To this end, the tools provided by this package provide a
user with most basic of descriptions of a galaxy's light, then allowing the
option to select any meaningful parameter by toggling a switch.  For most
parameters, such as aperture magnitudes, the switch is simple and
automatic.  For more complicated parameters, such as a profile fit or an
asymptotic magnitude, the switch is understandably more sophisticated and
needing more explanation to the user for accurate use.

\subsection{Basic Steps}

This paper is divided into five sections describing the major components of
the reduction package: 1) sky determination, 2) 2D profile fitting, 3)
aperture photometry, 4) extraction of 1D parameters from surface brightness
profiles and 5) extracting total magnitudes.  Each section contains
examples of the reduction of galaxy images from the 2MASS archive.

\subsection{Quick Start}

The fastest way to introduce the techniques and tools used in our package
is to walk through the analysis of a several different types of galaxy
images.  A more non-linear reader can refer to the Appendix for a listing
of the major tools.  A script titled $profile$ is included which outlines
the usage of the routines described below.  For a majority of galaxy
images, this script will produce a usable surface brightness profile, and
this script forms the core of the client/server version of this package
(see \S 5).  But, a sharper understanding of the data requires more
interaction with the techniques, the user is encouraged to run through the
examples given in the package.

To illustrate our tools, we have selected 2MASS $J$ images of several
galaxies found in the Revised Shapley-Ames catalog with the characteristics
of smooth elliptical shape (NGC 3193), disk shape (IC 5271), spiral
features (NGC 157) and low in surface brightness/low contrast to sky (NGC
2082).  The analysis procedure for each galaxy is divided into
five basic parts; 1) sky determination, 2) cleaning, 3) ellipse fitting, 4)
aperture photometry and 5) profile fitting.

Before starting it is assumed that the data frame has be initially
processed for flatfielding, dark subtraction and masking of chip defects.
Small defects, such as cosmic rays, are cleaned by the analysis routines.
But the errors are always reduced if known features are removed before
analysis.  The following routines work on poorly flattened data (e.g.
gradients or large-scale features), and will signal the poorness by the
resulting errors, but the removal of large-scale flattening problems
requires more interaction then acceptable for this package and remains the
responsibility of the user.

\subsection{Sky Determination}

Any galaxy photometry analysis process begins with an estimate of the
image's sky value.  While this is not critical for isophote fitting, it is
key for actually finding targets, cleaning the frame of stars and smaller
galaxies, plus determination of the photometry zeropoints.  Accurate sky
determination will, in the end, serve as the final limit to the quality of
surface photometry data since a majority of a galaxy's luminosity
distribution is near the sky value.  For this reason, sky determination has
probably received as much attention in astronomical data literature as any
other comparable reduction or analysis problem.

The difficulty in sky determination ranges from too few photons to know the
behavior of the instrumental response (e.g., high energy data) to a high
temporal varying flux of sky photons that overwhelms the galaxy signal
(e.g., near-IR data).  Surface fitting, drift scans, sky flats and super
flats are all procedures used to minimize the sky contribution to the noise
levels of the final data.  Several clever, but not technically challenging,
algorithms were included in the NOAO IRAF system to handle time averaged
flats and data, median co-adding and cosmic ray subtraction.  In the end,
improved CCD quality lowered the demands of sky subtraction as the
production of linear, good charge transfer and uniform sensitivity chips
replaced the earlier generations and their wildly inaccurate backgrounds.

\begin{figure}
\centering
\includegraphics[scale=0.40]{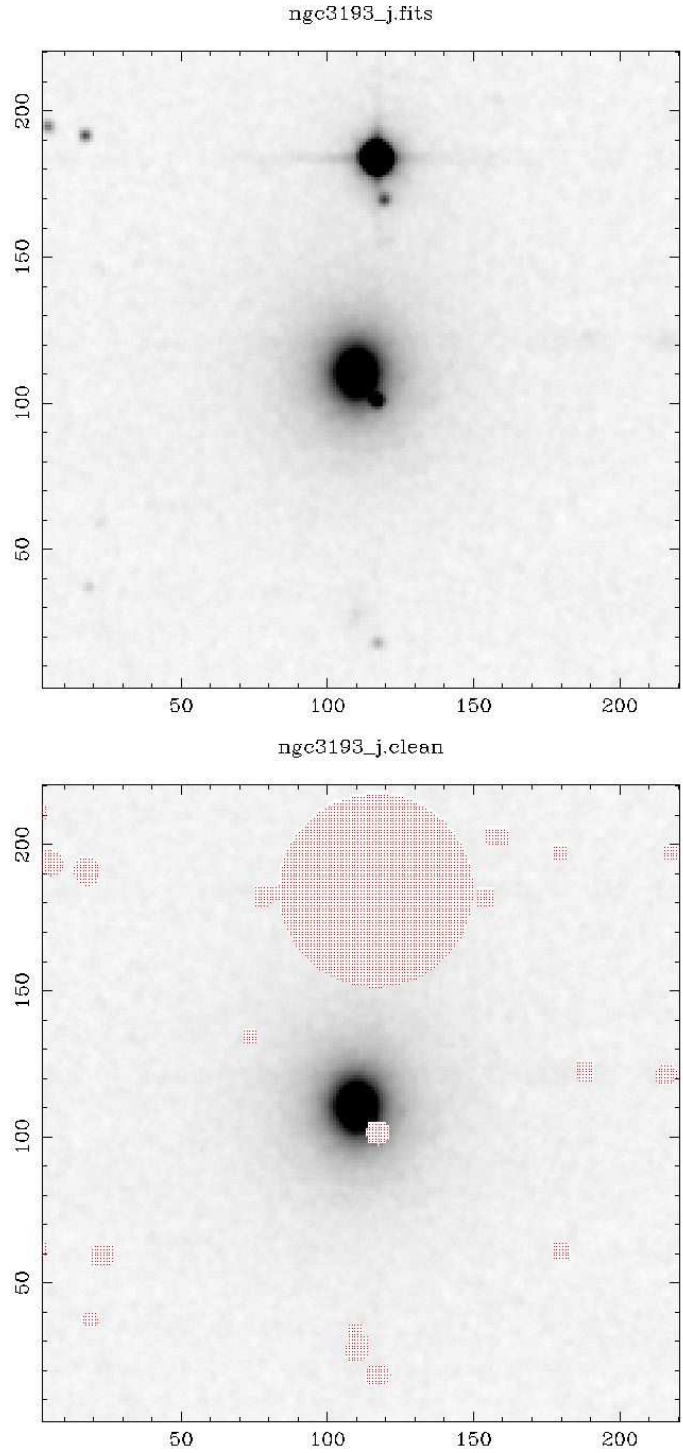}
\caption{The raw and cleaned $J$ frames for NGC 3193, an
elliptical selected from the RSA sample (Schombert
2007).  Note the proper cleaning of contaminating stars,
even a object near the galaxy core.
}
\end{figure}

For a cosmetically smooth image, an efficient, but crude, sky fit is one
that simply examines the border of the frame and does an iterated average,
clipping pixels more than 4$\sigma$ from the mean.  A border sky fit is
often sufficient to find the starting center of the galaxy (for the ellipse
fitting routines), clean the frame of stars/galaxies external to the object
of interest (the ellipse fitting routines will clean along the isophotes,
see below) and provide a preliminary error estimate to the photometry.
This error estimate is preliminary in that the true limiting error in the
surface (and aperture) photometry of large galaxies is not the RMS of an
isophote, but how well the sky value is know.  Once the number of pixels
involved in a calculation (be it an isophote or an aperture) becomes large
(greater than 50 for typical readout noises), then the error is dominated
by the precision of the sky value.

The disadvantage to a border sky fit is the occasional inconvenient
occurrence of stars or bright galaxies on the edge of the frame.  An
iterated mean calculation will remove small objects.  And large objects
will be signaled with large $\sigma$'s in an iterative mean search.  In an
automated procedure, more than likely, the task will have to halt and
request human intervention to find a starting sky value.

After years of experimentation, the method of choice for accurate sky
determination for extended galaxies is to evaluate sky boxes.  This is a
procedure where boxes of a set sized are placed semi-randomly (semi in the
sense of avoiding stars and other galaxies) in the frame.  An algorithm
calculates an iterative mean and $\sigma$ for each box.  These means (and
$\sigma$'s) are then super-summed to find the value of the sky as the mean
of the means (and likewise, the error on the sky value
is the $\sigma$ on this mean).

From an analysis point of view, there are several advantages to this
technique.  One is that each box exists as a measurement independent of the
other boxes.  Thus, the statistical comparison of the boxes is a real
measure of the quality of the sky determination for the frame in terms of
its accuracy and any gross trends with the frame.  Another advantage is
that contaminating objects are relatively easy to avoid (visual choice of
sky boxes) or to sort by the higher $\sigma$ per box.  Lastly, sky boxes
are the easiest method of finding regions for sky determination outside the
galaxy itself, particularly where an irregular object may fill a majority
of the data frame.

\begin{figure}
\centering
\includegraphics[scale=0.55]{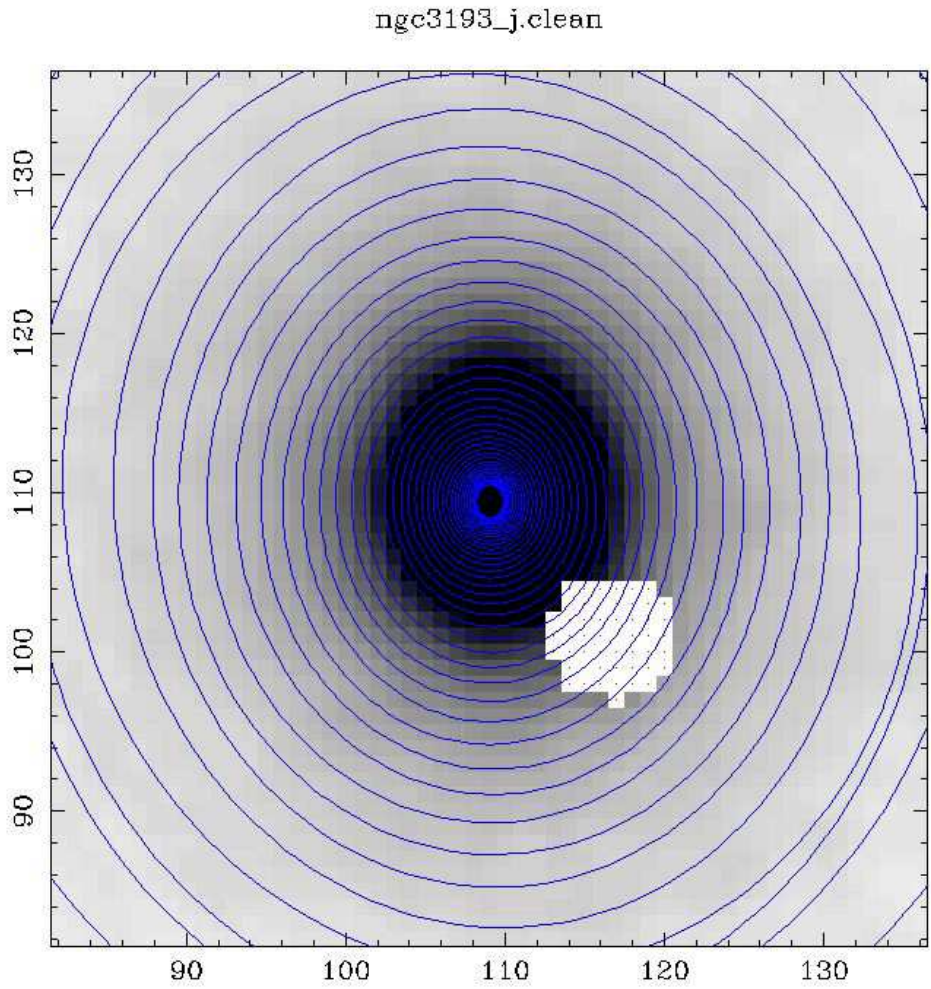}
\caption{The resulting ellipse fits to NGC 3193's core
region.  While the automatic masking of the
contaminating star is not perfect, it is sufficient to
maintain a high quality fit.
}
\end{figure}

The most difficult decision in sky determination by boxes is, of course,
where to place the boxes.  When done visually, the user selects region
(usually with a cursor) that are free of stars and sufficiently far away
from the target galaxy to be clear of its envelope light.  For an automated
process, the procedure returns for a final sky estimate after the
ellipse fitting process is completed and when all the stars/galaxies are
cleaned (set to values of not-a-number, NaN).  Then, the outer edge of the
large galaxy is determined and an iterative analysis of sky boxes outside
this radius is used to determine the true sky and, most importantly, the
variation on the mean of those boxes as a measure of how well the sky is
known.  This procedure is the role of {\it sky\_box}, see the Appendix for
a more detailed description of its options.

\subsection{Ellipse Fitting}

Reduction of a 2D image into a 1D run of intensity versus radius in a
galaxy assumes some shape to the isophote.  Very early work on galaxies
used circles since the data was obtained through circular apertures in
photoelectric photometers.  For early type galaxies, the ellipse is the
shape that most closely follows the shape of the isophotes.  This would
confirm that the luminosity being traced by an isophote is due to stellar
mass, which follow elliptical orbits (Kepler's 1st law).  As one moves to
along the Hubble sequence to later type galaxies, the approximation of an
ellipse to the isophotes begins to break down due to recent star formation
producing enhancements in luminosity density at semi-random positions.
However, no consistent shape describes the isophotes of irregular galaxies,
so an ellipse is the best shape, to first order, and provides a common
baseline for comparison to more regular galaxies.

\begin{figure}
\centering
\includegraphics[scale=0.40]{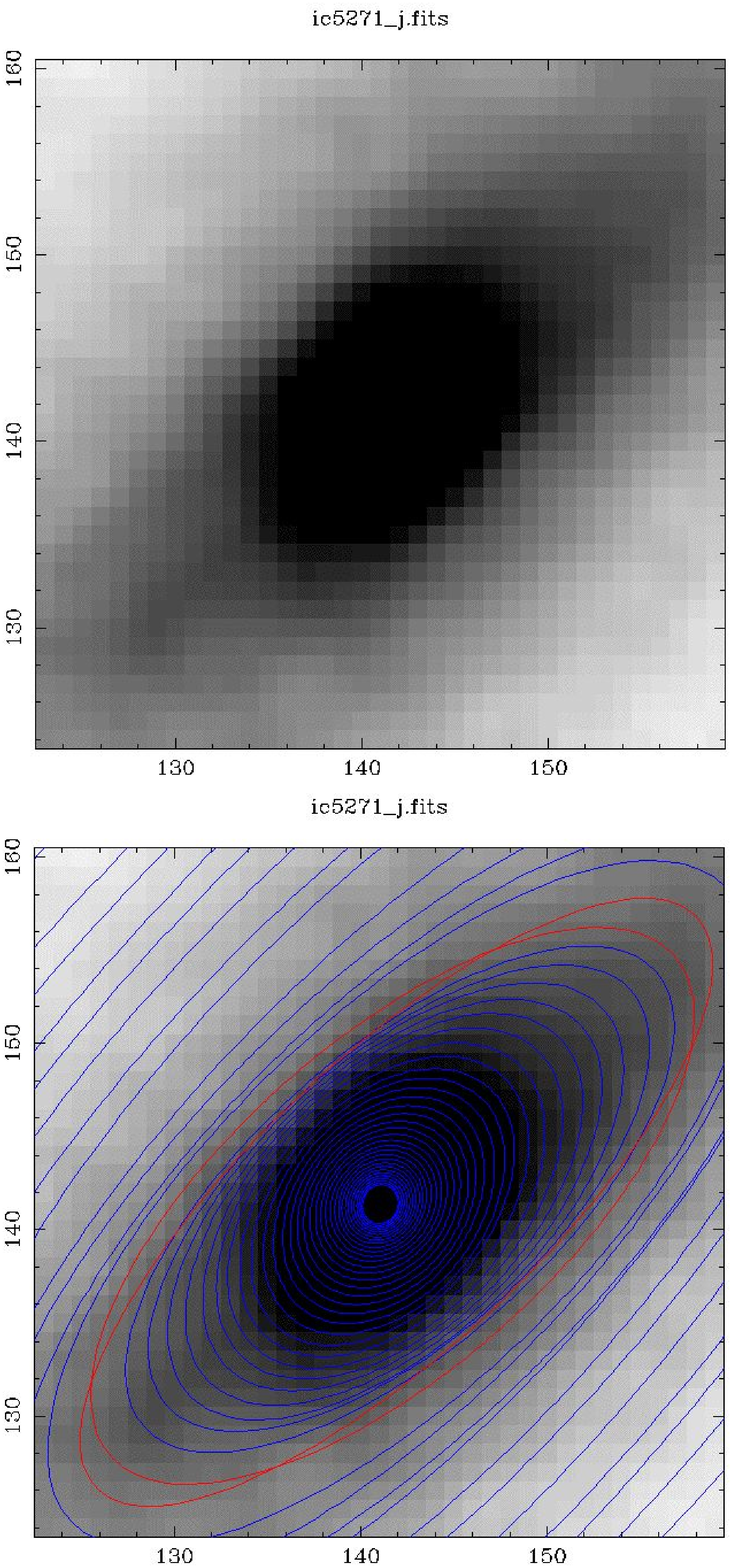}
\caption{A high contrast zoom-in of the 2MASS $J$ image of IC 5271, a Sb(rs) galaxy
selected from the RSA catalog.  Typical of the isophotes for a disk/bulge galaxy,
there is a cross over point as one transitions from a more spherical bulge to a
flattened disk.  While this is flagged by the reduction software, it is
astrophysically real and signals the lens morphology often seen in surface
brightness profiles of disk galaxies.
}
\end{figure}

Fitting a best ellipse to a set intensity values in a 2D image is a
relatively straight forward technique that has been pioneered by Cawson
\etal (1987) and refined by Jedrzejewski (1987) (see also an excellent
review by Jensen \& Jorgensen 1999).  The core routine from these
techniques (PROF) was eventually adopted by STSDAS IRAF (i.e. ELLIPSE).
The primary fitting routine in this package follows the same techniques (in
fact, uses much of the identical FORTRAN code from the original GASP
package of Cawson) with some notable additions.

These codes start with an estimated x-y center, position angle and
eccentricity to sample the pixel data around the given ellipse.  The
variation in intensity values around the ellipse can be expressed as a
Fourier series with small second order terms.  Then, an iterative
least-squares procedure adjusts the ellipse parameters searching for a best
fit, i.e. minimized coefficients.  There are several halting factors, such
as maximum number of iterations or minimal change in the coefficients,
which then moves the ellipse outward for another round of iterations.  Once
a stopping condition is met (edge of the frame or sufficiently small change
in the isophote intensity), the routine ends.  A side benefit to above
procedure is that the cos(4$\theta$) components to each isophote fit are
easily extracted, which provides a direct measure of the geometry of the
isophote (i.e. boxy versus disk-like, Jedrzejewski 1987).

One new addition, from the original routines, is the ability to clean (i.e.
mask) pixels along an isophote.  Basically, this routine first allows a few
iterations to determine a mean intensity and RMS around the ellipse.  Any
pixels above (or below) a multiple of the RMS (i.e. 3$\sigma$) are set to
not-a-number (NaN) and ignored by further processing.  Due to the fact that
all objects, stars and galaxies, have faint wings, a growth factor is
applied to the masked regions.  While this process is efficient in
early-type galaxies with well defined isophotes, it may be incorrect in
late-type galaxies with bumpy spiral arms and HII regions.  The fitting
will be smoother, but the resulting photometry will be underestimated.
This process can be controlled early in the analysis pipeline by the user
with an initial guess of the galaxy's Hubble type.  Also, the erased pixels
are only temporary stored until an adequate fit is found.  Once a
satisfactory ellipse is encountered, only then are the pixels masked for
later ellipse fitting.  The masked data is written to disk at the end of
the routine as a record of the cleaning.  The ellipse fitting is the
function of {\it efit} as described in the Appendix.

For early-type galaxies, lacking any irregular features, the cleaning process
is highly efficient.  The pipeline first identifies the galaxy and its
approximate size by moment analysis.  It then cleans off stars/galaxies
outside the primary galaxy by moment identification and radius growth for
masking.  Stars/galaxies inside the primary galaxy are removed by the ellipse
fitting routine.  The resulting ellipses are inspected for crossover
(isophotes that crossover are assumed to be due to errors or embedded
stars/galaxies and removed by averaging nearby ellipse isophotes, this is
not true for disk galaxies).  The smoothed ellipses are used by a more
robust cleaning algorithm and the whole ellipse fitting process is
repeated on the cleaned frame.

\begin{figure}
\centering
\includegraphics[scale=0.45]{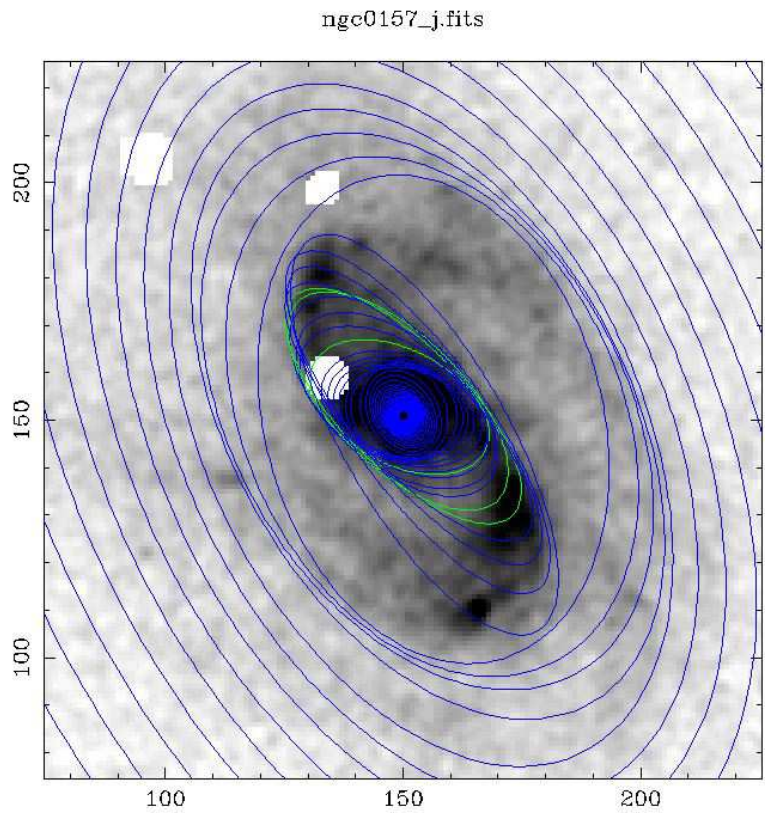}
\caption{A high contrast zoom-in of the 2MASS $J$ image of NGC 157, a
face-on Sc(s) galaxy selected from the RSA catalog.  Similar to IC 5271,
there are several crossover points in the fitted ellipses.  The fitting
program does a good job of following the spiral arms in the inner regions,
then a large jump from bulge to disk region.
}
\end{figure}

An example of the analysis of an elliptical is found in Figure 1, a 2MASS $J$
image of NGC 3193.  The top panel is the raw 2MASS image, the bottom panel
is the resulting cleaned image output at the end of the reduction process.
The cleaning process efficiently removed all the stars on the frame,
including the brighter object on the northern edge of the frame and its
diffraction spikes.  The star closest to the galaxy core is a problem in
two arenas.  The first is in the calculation of ellipse, as the inner star
would drag the calculated moments off center.  The isophote erasing routine
has handled this as can be seen in Figure 2, where the fitting ellipses are
shown and are not deflected by the erased star.  Second, is that calculated
total magnitudes would either be over estimated (if the star is not masked)
or under estimated (if the star is masked and the galaxy light from those
pixels is not replaced).  This problem will be discussed in \S 2.6.

As one goes towards later type galaxies, there is an increase in the
non-elliptical nature to their isophotes and an increase in luminosity
density enhancements (HII regions, stellar clusters, spiral features) which
are legitimate components to the galaxy's light distribution and should not
be cleaned.  The user can specify the galaxy type and the cleaning
restrictions will be tightened (only to stellar objects and at a higher
cleaning threshold) plus the restrictions on overlapping ellipses is
loosened (e.g. the transition from a round bulge to a flat disk).  Most
importantly, while some galaxy features are cleaned for the sake of a
harmonious ellipse fit, those pixels need to be filled for later aperture
photometry.

\begin{figure}
\centering
\includegraphics[scale=0.40]{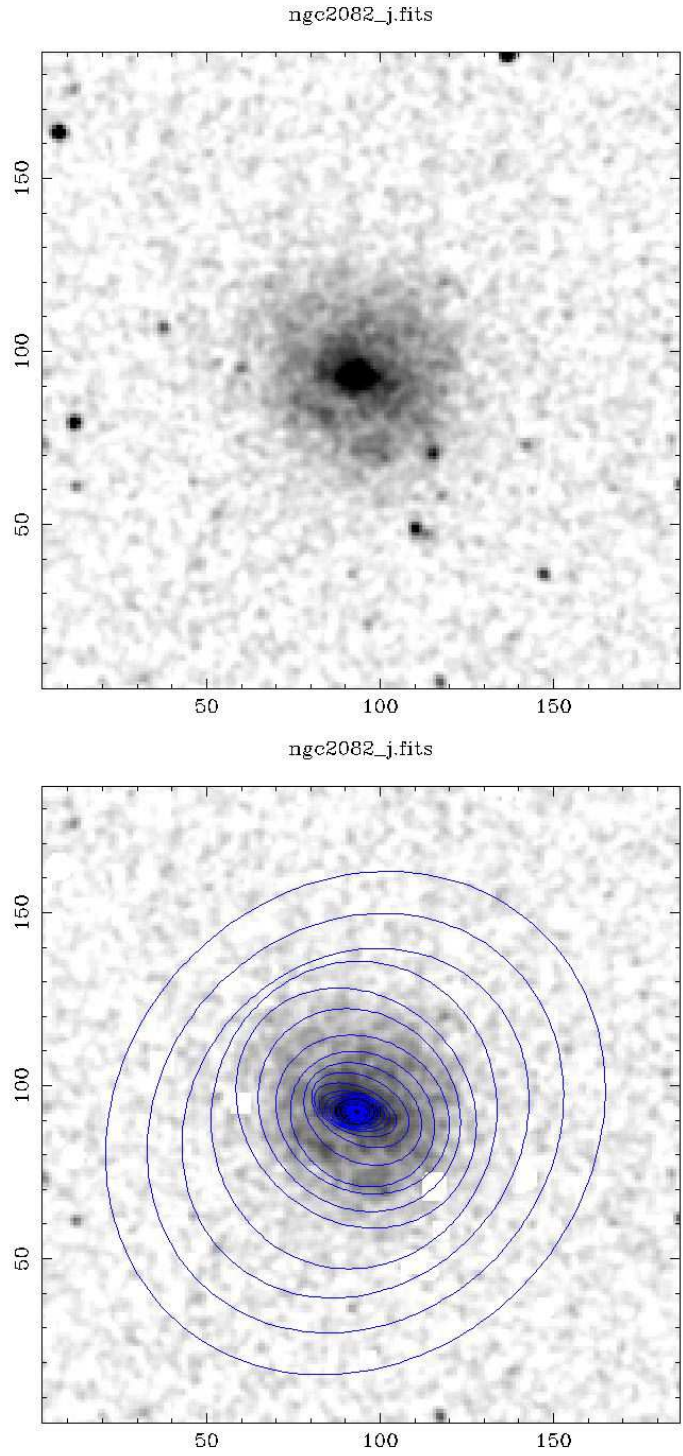}
\caption{A high contrast subimage of the 2MASS $J$ image of NGC
2082, a LSB disk galaxy.  Note, that the ellipse fitting routine
expanded the annulus size to increase the S/N. 
}
\end{figure}

An example of this behavior can be found in Figure 3, the $J$ image of IC
5271.  The red ellipses indicate isophote fits that crossover.  While
flagged as an error, this is in fact the real behavior of the isophotes as
one transitions from bulge to disk.  The resulting intensities are probably
overestimated due to the crossover effect, but this error will be minor
compared to the errors that would result from an off-center or overly round
ellipse.

The quality of the fitting procedure can be judged by the behavior of the
ellipse parameters such as eccentricity, position angle and center.  If
there are large jumps in any of the parameters that determine shape, then
this may signal a feature in the galaxy that needs to be cleaned (a buried
star for example).  Slightly less abrupt changes may signal an astrophysically
interesting features, such as a bar or lens morphology.  Under the
assumption that the isophotes of a typical galaxy are a smooth function
with radius, the ellipse fitting algorithm checks for ellipse parameters
that indicate a crossing of the isophotal lines.  These ellipses are
smoothed and flagged (the mean of the inner and outer ellipse parameters is
used).  In certain scenarios, crossing isophotes are to be expected, for
example the transition region from a bulge to a disk (see Figure 2), and
the smoothing criteria is relaxed.  This is the function of $prf\_smooth$ as
described in the Appendix.

\begin{figure}
\centering
\includegraphics[scale=0.45]{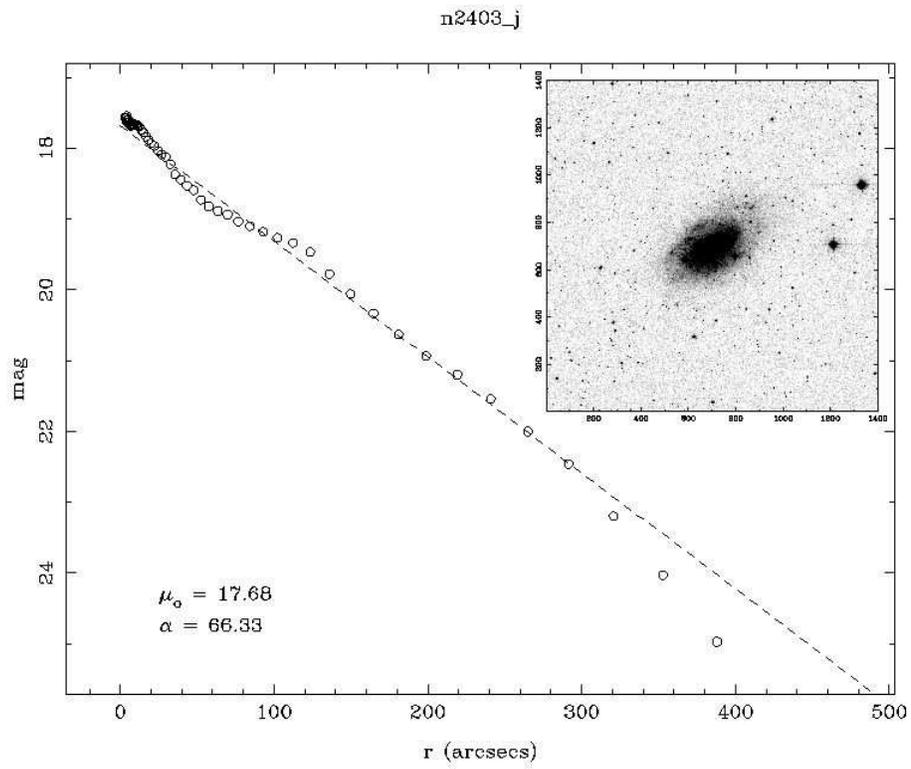}
\caption{A final surface brightness profile for NGC
2403, an Sc(s) galaxy.  The dotted line is a bi-weighted linear fit along
with the resulting fit parameters.
}
\end{figure}

An example of $prf\_smooth$'s corrections can be seen in Figure 4, the $J$
image of NGC 157.  Several interior ellipses display erratic behavior, but
$prf\_smooth$ took the mean average of nearby ellipses (in green) to
produce a more rational fit.  The resulting intensities were also more
stable, although the RMS is going to be highly than the
typical isophotes found in an elliptical.

An example of a LSB galaxy fit is found in Figure 5.  The ellipse fitting
routine, recognizing that the target is low in contrast with respect to
sky, widened the annulus for collecting pixel values.  This increases the
S/N at some loss of spatial resolution.  Since resolution is usually not
important in a galaxy's halo region, this is an acceptable trade off.  LSB
galaxies are susceptible to fitting instability, the fitting routines are
tightened against rapid changes in eccentricity and centering to prevent
this behavior.

LSB galaxies also demonstrate a key point in determining errors from
surface photometry.  There are two sources of error per isophote, the RMS
around the ellipse and the error in the sky value.  The RMS value is a
simple calculation using the difference between the mean and the individual
pixel values.  This RMS then reflects into an observable error as the
$\sqrt{N}$.  However, as the isophote intensity approaches the sky value,
the number of pixels increases and the error due to RMS becomes an
artificially low value.  In fact, at low intensities, the knowledge of the
sky value dominates and the error in the isophote is reflected by the sky
error (preferably as given by the $\sigma$ on the means of a large number
of sky boxes).

\subsection{Surface Photometry}

With a file of isophotal intensities versus radius in hand, it is a simple
step to producing a surface brightness profile for the galaxy.  There are a
few tools are in the package to examine the quality of the ellipse fitting
(e.g. $prf\_edit$, an interactive comparison of the image and the
ellipses).  At the very least, a quick visual inspection of the ellipses
seems required as a bad mismatch leads to strongly biased results (see
cautionary tale in \S 4).  A user can either step through a directory of
data files (e.g. using the $probe$ tool) or a user can automatically
produce a group of GIF images with a corresponding HTML page, then use a
browser to skim through a large number of files.  Calibration from image
data numbers (DN) to fluxes (or magnitudes) is usually obtained through
standard stars with corrections for airmass and instrumental absorption.
If these values are in the FITS headers, then they are automatically added
to the object's XML file.  Additional corrections for galactic absorption,
k-corrections and surface brightness dimming are well documented in the
literature and can be assigned automatically by grabbing XML data from NED.
A chosen cosmology converts radius in pixel units into astrophysically
meaningful values of kiloparsecs.  A Python command line script ($cosmo$)
based on Ned Wright's cosmology calculator is included in the package   All
these values can be added to the XML file for automatic incorporation to
the analysis programs.  If they don't exists, then instrumental mags will be
used, which can easily be converted to real units later on.

\begin{figure}
\centering
\includegraphics[scale=0.45]{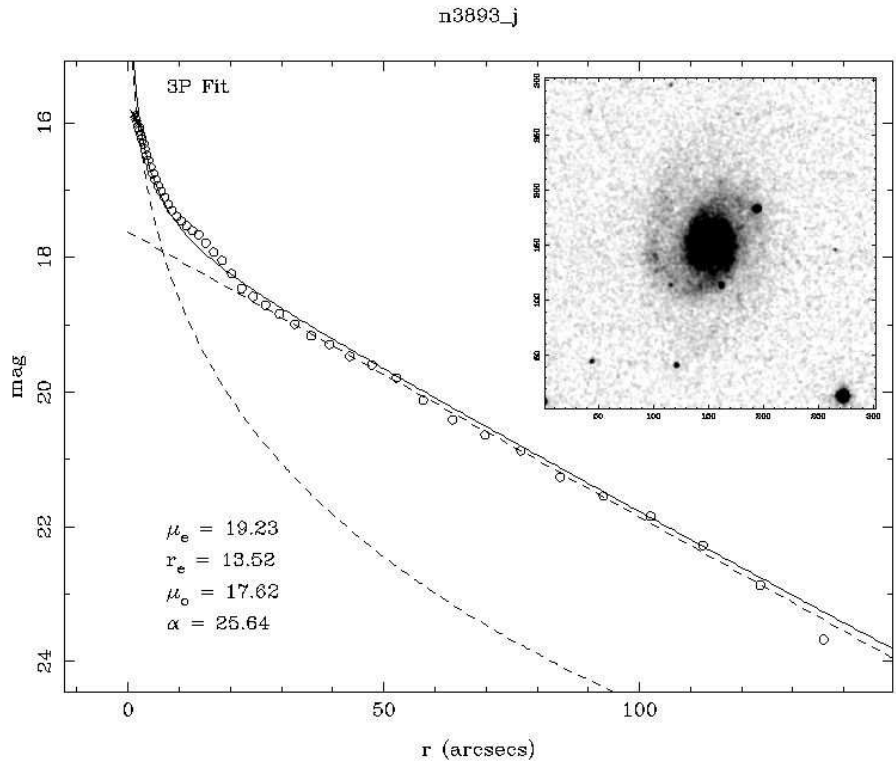}
\caption{A final surface photometry profile for NGC
3893, an Sc(s) galaxy.  The dotted lines are exponential and r$^{1/4}$
fits to disk and bulge.  The solid line is the addition
of the two curves.
}
\end{figure}

Analysis of a 1D surface brightness profile (the job for the $bdd$ tool)
depends on the scientific goals of the user.  For example, early-type
galaxies are typically fit to a de Vaucouleurs r$^{1/4}$ curve to extract a
scale length (effective radius) and characteristic surface brightness
(effective surface brightness).  Irregular and dwarf galaxies are well fit
by exponential profiles which provide a disk scale length and central
surface brightness.  Disk galaxies can be fit with a combination of bulge
and disk fits, to extract B/D ratios and disk scale lengths.  

Due to this combination of $r^{1/4}$ and exponential curves for large bulge
spirals, it is computationally impossible to correctly determine which
function, or combination of functions, best fits a particular galaxy's
profile.  In the past, one would examine the 2D image of the galaxy and
obvious disk-like galaxies would be fit to $r^{1/4}$ plus exponential.
Objects with elliptical appearance were fit to a strict $r^{1/4}$ shape.
This produces a problem for large bulge S0's which are difficult to detect
visually unless nearly edge-on.

The simplest solution to this problem, using only the 1D surface
photometry, is to examine the profiles in a plot of mag arcsecs$^2$ versus
linear radius.  With this plot, exponential disks appear as straight lines,
see Figure 6 as an example of a pure disk in NGC 2403.  Bulge plus disk
components are also straight forward in this mag/linear radius space, see
Figure 7 a good example of a bulge plus disk fit in NGC 3983.  If a profile
displays too much curvature, with no clear linear disk portion, then it is
a good candidate for a pure $r^{1/4}$ fit (see Figure 8, NGC 3193).  This
option is easily checked by plotting the profile in mag arcsecs$^2$ versus
r$^{1/4}$ space as shown in Figure 8.  Most r$^{1/4}$ profiles only have a
linear region in the middle of the surface brightness profile, typically
with a flattened core and fall-off at large radii (see Schombert 1987).

\begin{figure}
\centering
\includegraphics[scale=0.45]{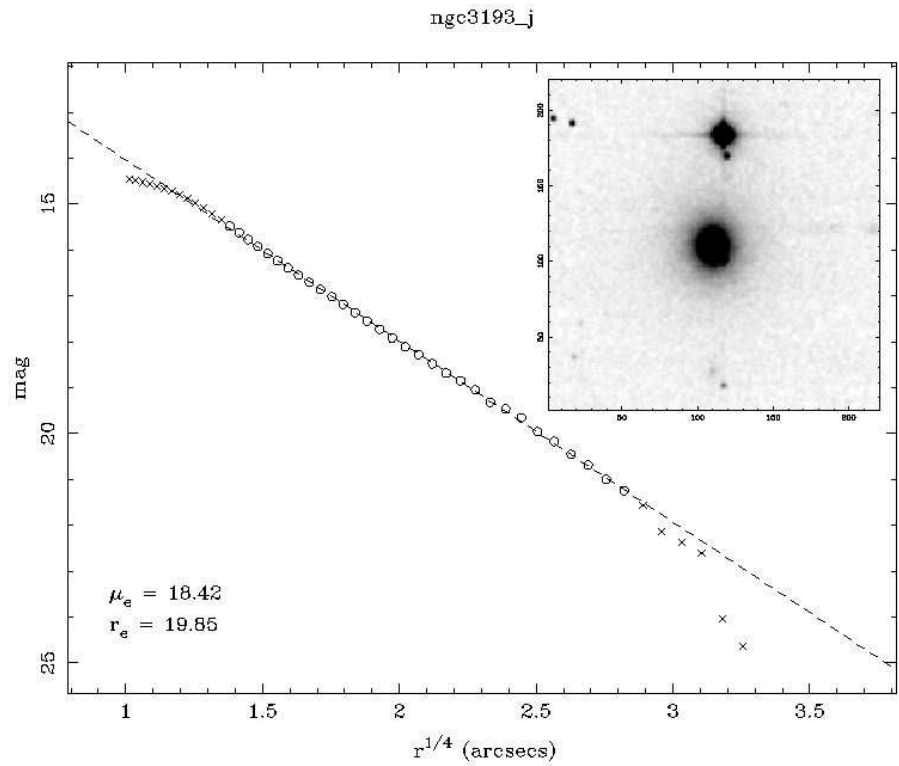}
\caption{A final surface photometry profile for the
elliptical NGC 3193.  A pure r$^{1/4}$ fit is shown.
}
\end{figure}

The Seric function is also popular for fitting surface brightness profiles
(Graham \& Driver 2005), although not currently supported by this package,
any fitting function is easy to add to the reduction routines as the core
search routine is a grid search $\chi^2$ minimization technique.  However,
there are issues with surface photometric data where the inner regions have
the highest S/N but the outer regions better define a galaxy's structure
(Schombert \& Bothun 1987).  With user guidance, this grid search
works well for any user defined function.  Also, since there are a
sufficient number of packages for fitting 1D data in the community, 
this package only provides a simple graphic plotting function.  More
sophisticated analysis needs guidance by the user, but this package
provides the framework for just such additions.  

\subsection{Aperture Photometry}

Often the scientific goal of a galaxy project is to extract a total
luminosity for the system (and colors for multiple filters).  For small
galaxies, a metric aperture or isophotal magnitude is suitable for
comparison to other samples (certainly the dominate source of error will
not be the aperture size).  However, for galaxies with large angular size
(i.e. many pixels), their very size makes total luminosity determination
problematic.

\begin{figure}
\centering
\includegraphics[scale=0.45]{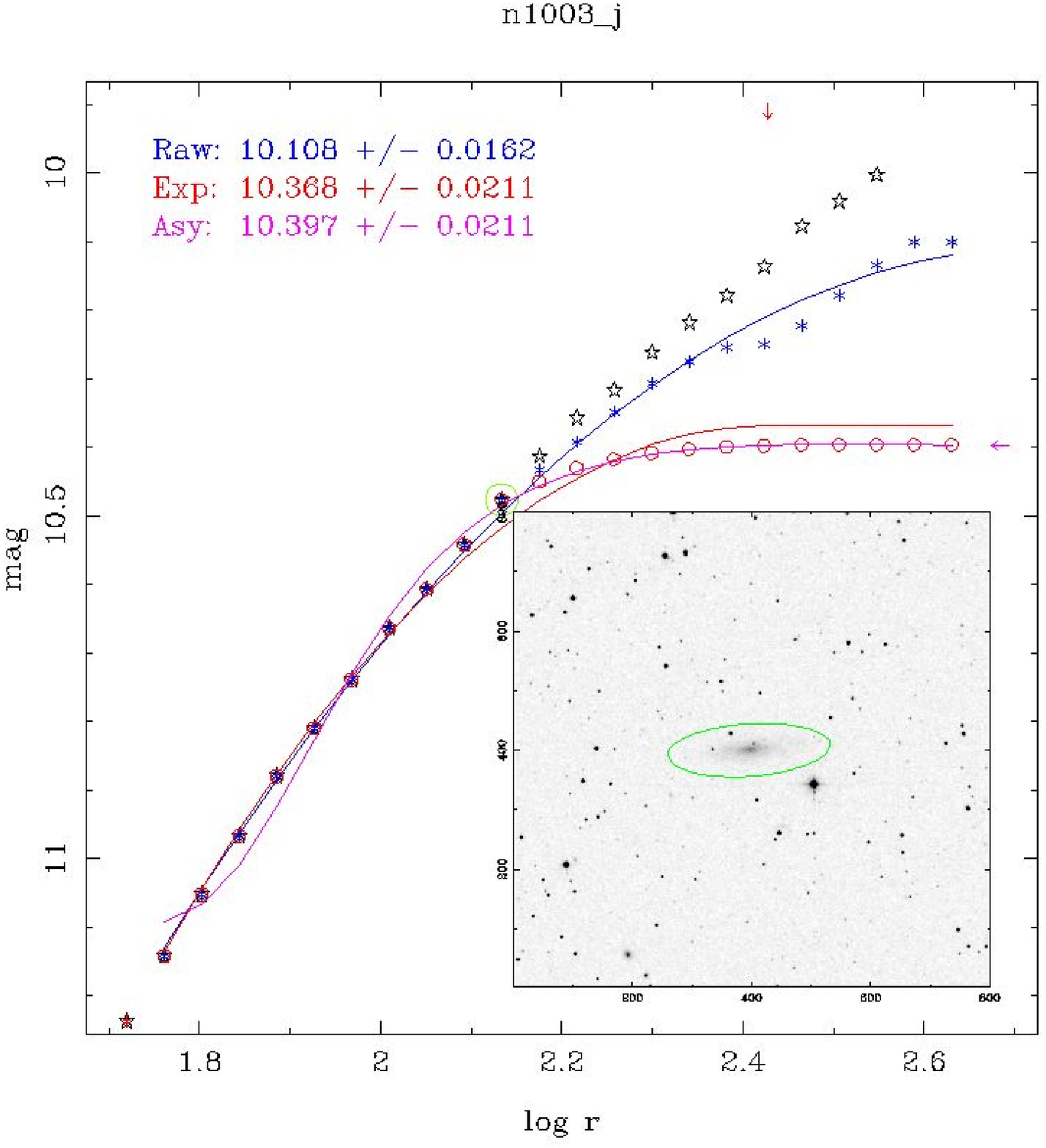}
\caption{Elliptical aperture photometry of LSB galaxy,
NGC 1003.  Starred data is the raw intensities, asterisks
are apertures determined from ellipse fitted isophotes,
circles are interpolation from the surface brightness
fits.  Blue and orange lines are 2nd order polynomial fits,
the pink line is a fit using rational functions.
}
\end{figure}

Natively, one would think that a glut of pixels would make the problem of
determining a galaxies luminosity easier, not more difficult.  However, the
problem here arises with the question of where does the galaxy stop?  Or,
even if you guess an outer radius, does your data contain all the galaxy's
light?  The solution proposed by de Vaucouleurs' decades ago is to use a
curve of growth (de Vaucouleurs 1977).  Almost all galaxies follow a
particular luminosity distribution such that the total light of a galaxy
can be estimated by using a standard growth curve to estimate the amount of
light outside your largest aperture.  For a vast majority of galaxies,
selecting either an exponential or r$^{1/4}$ curve of growth is sufficient
to adequately describe their total luminosities (Burstein \etal 1987).
However, for modern large scale CCD imaging, the entire galaxy can easily
fit onto a single frame and there is no need for a curve of growth as all
the data exists in the frame.

With adequate S/N, it would seem to be a simple task to place a
large aperture around the galaxy and sum the total amount of light (minus
the sky contribution).  However, in practice, a galaxy's luminosity
distribution decreases as one goes to larger radii, when means the sky
contribution (and, thus, error) increases.  In most cases, larger and
larger apertures simply introduce more sky noise (plus faint stars and
other galaxies).  And, to further complicate matters, the breakover point
in the optical and near-IR, where the galaxy light is stronger than the sky
contribution will not contain a majority of the galaxy's light.  So the
choice of a safe, inner radius will underestimate the total light.

The procedure selected in this package, after some numerical
experimentation, is to plot the aperture luminosity as a function of radius
and attempt to determine a solution to an asymptotic limit of the galaxy's
light.  This procedure begins by summing the pixel intensities inside the
various ellipses determined by $efit$.  For small radii, a partial pixel
algorithm is used to determine aperture luminosity (using the surveyors
technique to determine each pixel's contribution to the aperture).  At
larger radii, a simply sum of the pixels, and the number used, is output.
In addition, the intensity of the annulus based on the ellipse isophote and
one based on the fit to the surface photometric profile are also outputted
at these radii (see below).

Note that a correct aperture luminosity calculation requires that both a
ellipse fit and a 1D fit to the resulting surface photometry has be made.
The ellipse fit information is required as these ellipses will define the
apertures, and masked pixels are filled with intensities given by the
closest ellipse.  A surface photometric fit allows the aperture routine to
use a simple fit to the outer regions as a quick method to converge the
curve of growth.

Once the aperture luminosities are calculated, there are two additional
challenges to this procedure. The first is that an asymptotic fit is a
difficult calculation to make as the smallest errors at large radii reflect
into large errors for the fit.  Two possible solutions are used to solve
this dilemma.  The first solution is to fit a 2nd or 3rd order polynomial
to the outer radii in a luminosity versus radius plot.  Most importantly
for this fit, the error assigned the outer data points is the error on the
knowledge of the sky, i.e. the RMS of the mean of the sky boxes.  This is
the dominant source of error in large apertures and the use of this error
value results in a fast convergence for the asymptotic fit.  The resulting
values from the fit will be the total magnitude and total isophotal size,
determined from the point where the fit has a slope of zero.  A second
solution is to use an obscure technique involving rational functions.  A
rational function is the ratio of two polynomial functions of the form

$$ f(x) = {{a_nx^n+a_{n-1}x^{n-1}+ ... + a_2x^2+a_1x+a_0} \over 
{b_mx^m+b_{m-1}x^{m-1}+ ... + b_2x^2+b_1x+1}} $$

\noindent where $n$ and $m$ are the degree of the function.  Rational
functions have a wide range in shape and have better interpolating
properties than polynomial functions, particularly suited for fits to data
where an asymptotic behavior is expected.  A disadvantage is that rational
functions are non-linear and, when unconstrained, produce vertical
asymptotes due to roots in the denominator polynomial.  A small amount of
experimentation found that the best rational function for aperture
luminosities is the quadratic/quadratic form, meaning a degree of 2 in the
numerator and denominator.  This is the simplest rational function and has
the advantage that the asymptotic magnitude is simply $a_2/b_2$, although
is best evaluated at some radii in the halo of the galaxy under study.

Usually the aperture luminosity values will not converge at the outer edges
of a galaxy.  This is the second challenge to aperture photometry, correct
determination of the luminosity due to the faint galaxy halo.  This is
where the surface photometry profile comes in handy.  Contained in that
data is the relationship between isophotal luminosity and radius, using all
the pixels around the galaxy.  This is often a more accurate number than
attempting to determine the integrated luminosity in an annulus at the same
radius.  This information can be used to constraint the curve of growth in
two ways.  One, we can use the actual surface brightness intensities and
convert them to a luminosity for each annulus at large radii. Then, this
value can be compared to the aperture value and a user (or script) can flag
where the two begin to radically deviate.  Often even the isophotal
intensities will vary at large radii and, thus, a second, more stable
method is to make a linear fit of an exponential, r$^{1/4}$ or combined
function to the outer radii and interpolate/extrapolate that fit to correct
the aperture numbers.

Figure 9 display the results for all three techniques for the galaxy NGC
1003.  The black symbols are the raw intensities summed from the image
file.  The blue symbols are the intensities determined from the surface
photometry.  The orange symbols are the intensities determined from the
fits to the surface photometric profile.  This was one of the worst case
scenarios due to the fact the original image is very LSB (in the near-IR
$J$ band).  Due to noise in the image and surface photometry, the outer
intensities grow out of proportional to the light visible in the greyscale
figure.  A fit to the raw data does not converge (blue line).  A 2nd order
fit to the profile fit (orange line) also fails to capture the asymptotic
nature.  The rational function fit (pink line) does converge to an accurate
value.  If similar types of galaxies are being analyzed, it is a simple
procedure to automate this process.

\section{Data files and XML}

In the past, when disk space was at a premium and I/O rates were slow,
astronomical data was stored in machine specific formats.  However, today
disk space is plentiful and file access times are similar to processing
times on most desktop systems.  Thus, a majority of simple astronomical
databases are stored in flat file format, also called plain text or ASCII
files (note: this is an interesting throwback to the original data methods
from the end of the 19th century, where information is stored as a system
of data and delimiters, such as spaces or commas).  The endproduct data
files for most packages rarely exceed a few kilobytes or a few hundred
lines.  The simplest access to these types of files is an editor such as vi
or emacs.  Sufficient documentation (i.e. header files) makes understanding
the data, and writing applications for further analysis these data files, a
relatively simple task.

However, there is a strong driver to migrate output files into XML format
for even the simplest data files.  Extensible Markup Language (XML) is a
W3C-recommended general-purpose markup language that is designed store and
process data.  The core of XML is the use of tags, similar to HTML tags
(i.e. \texttt{<tag>data</tag>}), to delineate data values and assign
attributes to those values (for example, units of measure).  XML is not
terse and, therefore, somewhat human-legible (see Figure 1 for XML example
of astronomical data).

\begin{figure}
\centering
\includegraphics[scale=0.40]{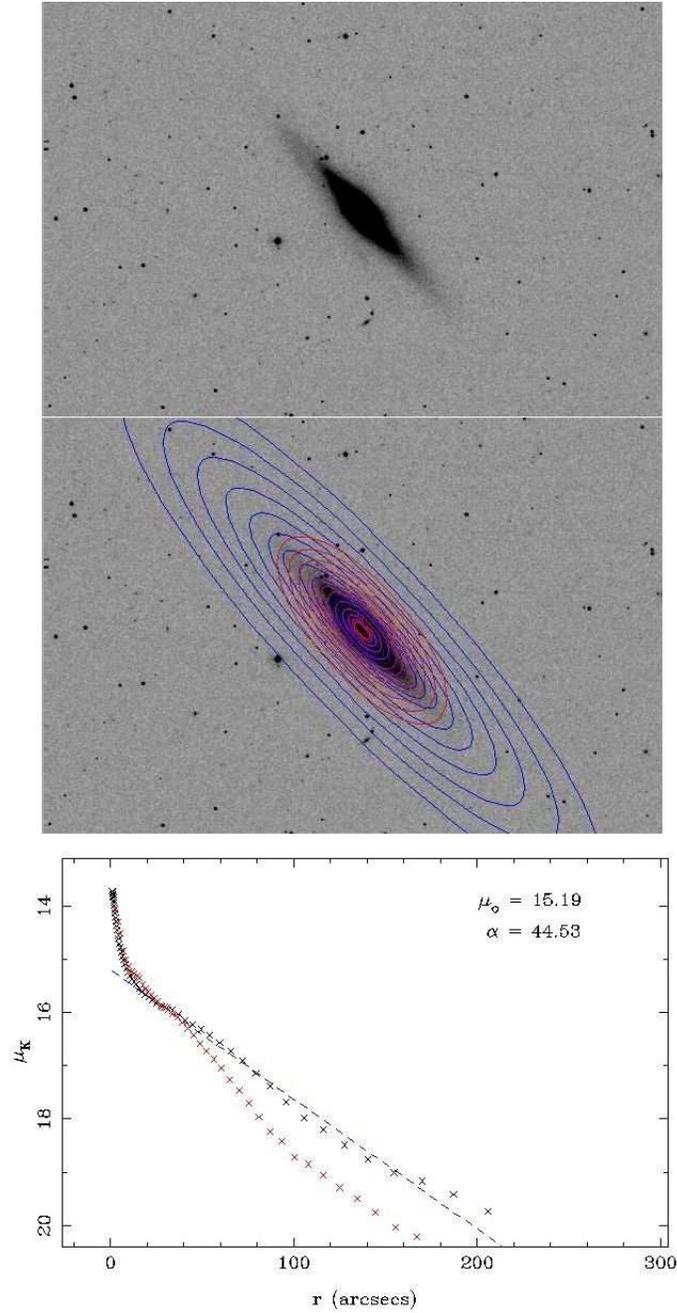}
\caption{The edge-on Sb galaxy, NGC 2683 selected from the 2MASS Large
Galaxy Atlas.  The top and middle panels display greyscale images from the
2MASS $J$ scans.  The red ellipses are fits from the 2MASS galaxy pipeline.
The blue ellipses are the resulting fits from this package.  The 2MASS fits
clearly fail to follow the flatter isophotes in the outer regions.  This
results in an underestimate of these isophote's intensity values, as seen
in the surface brightness profile in the bottom panel.
}
\end{figure}

XML has several key advantages over plain file formats.  For one, XML
format allows an endless amount of additional information to be stored in
each file that would not have fit into the standard data plus delimiter
style.  For example, calibrating data, such as redshift or photometric
zeropoint, can be stored in each file along with the raw data with very
little increase in file size overhead as the tags handle the separation.
There is no need to reserve space for these quantities nor is there any
problem adding future parameters to the XML format.  Using XML format puts
all the reduction data into a single file for compactness and, in addition,
since XML files are plain text files, there is no problem with machine to
machine transfer.  The reading of XML files is not a complication for
either compiled or interpreted languages.

A disadvantage to XML format is that it's clumsy to read.  However, there
exist a number of excellent XML editors on the market (for example,
http://www.oxygenxml.com).  These allow a GUI interface with an efficient
query system to interact with the XML files.   While many users would
prefer to interact with the raw data files in plain text form, in fact,
even a simple editor is GUI window into the bytes and bits of the actual
data on machine hardware.  A GUI XML editor is simply a more sophisticated
version of vi or emacs.

\begin{figure}
\centering
\includegraphics[scale=0.40]{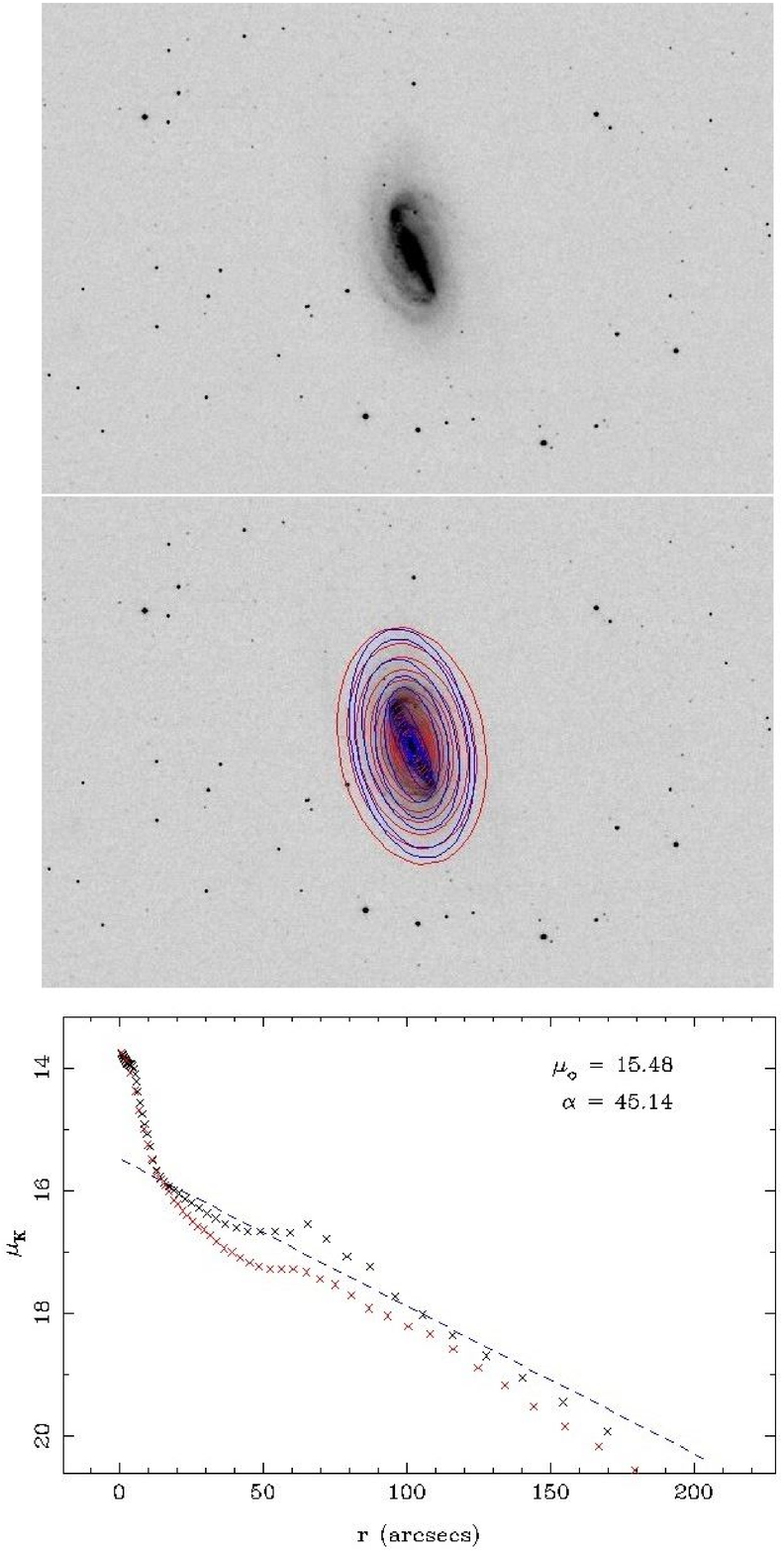}
\caption{The Sc(s) galaxy, NGC 2903 selected from the 2MASS Large Galaxy
Atlas.  The top and middle panels display greyscale images from the 2MASS
$J$ scans.  The red ellipses are fits from the 2MASS galaxy pipeline.  The
blue ellipses are the resulting fits from this package.  The 2MASS galaxy
pipeline failed to follow the isophote twists (i.e. changes in position
angle).  This results in an underestimate of these isophote's intensity
values, as seen in the surface brightness profile in the bottom panel.
}
\end{figure}

An additional reason to migrate to XML is a new power that XML data files
bring to data analysis.  Many interpreted languages (i.e.  Python and Perl)
have an {\it eval} or {\it exec} function, a method to convert XML data
into actual variables within the code at runtime (i.e.  dynamically typed).
This has a powerful aspect to analysis programs as one does not have to
worry about formats or the type of data entries, this is handled in the
code itself.  Dynamical typing introduces a high level of flexibility to
code.  In Python, one can convert XML data (using Python's own XML modules
to read the data) into lists that contain the variable name and value, then
transform these lists into actual code variables using an {\it exec}
command.  For example,

\begin{verbatim}
for var, value in xml_vars:
  exec(var+'='+value)
\end{verbatim}

\noindent produces a set of new variables in the running code.  And
Python's unique try/except processing traps missing variables without
aborting the routine.  For example, if the variable 'redshift' exists in
the XML data file then

\begin{verbatim}
try:
  distance=redshift/H_o
except:
  print 'redshift undefined'
  distance=std_distance
\end{verbatim}

\noindent This same try/except processing also traps overflows and other
security flaws that might be used by a malicious user attempting to
penetrate your server using the XML files.  Thus, XML brings a level of
security as well as enhancing your code.

Lastly, another advantage to XML format is the fact that all of the
reduction data (ellipse fitting, aperture photometry, calibration
information, surface photometry) can be combined into a single file, e.g.
galaxy\_name.xml, which can be interrogated by any analysis routine that
understands XML.  A simple switch at the end of the reduction process
integrates the data into an XML file for transport, or access by plotting
packages, etc.

\section{A Cautionary Tale, the 2MASS Large Galaxy Atlas}

If you have read this far, and are still awake, this section walks through
the reduction of part of the Revised Shapley-Ames sample (Schombert 2007)
taken from the 2MASS database that overlaps the 2MASS Large Galaxy Atlas
(Jarrett \etal 2003).  As a cautionary tale to the importance to doing
large galaxy photometry with care, we also offer in this section a
comparison of our technique with the results from an automated, but much
cruder reduction pipeline from the 2MASS project.

Allowing the ellipses to vary from isophote to isophote, not only in
eccentricity, but also in position angle and ellipse center, are critical
to obtaining an accurate description of a galaxy's luminosity profile.
Shown in Figures 10, 11 and 12 are examples of $J$ images extracted from the
2MASS archives that were part of the Large Galaxy Atlas (Jarrett \etal
2003).  In each case, the 2MASS pipeline calculates a luminosity profile
based on the isophotes around an ellipse from the mean moments of the whole
galaxy.  Thus, the fitted ellipses do not change in axial ratio or position
angle and, for most spirals, this technique will result in an ellipse that is
too flat in the core and, often, too round in the halo regions.  If the
galaxy has a bar, this technique will also underestimate the bar contribution,
spreading its light into larger radii ellipses.

\begin{figure}
\centering
\includegraphics[scale=0.40]{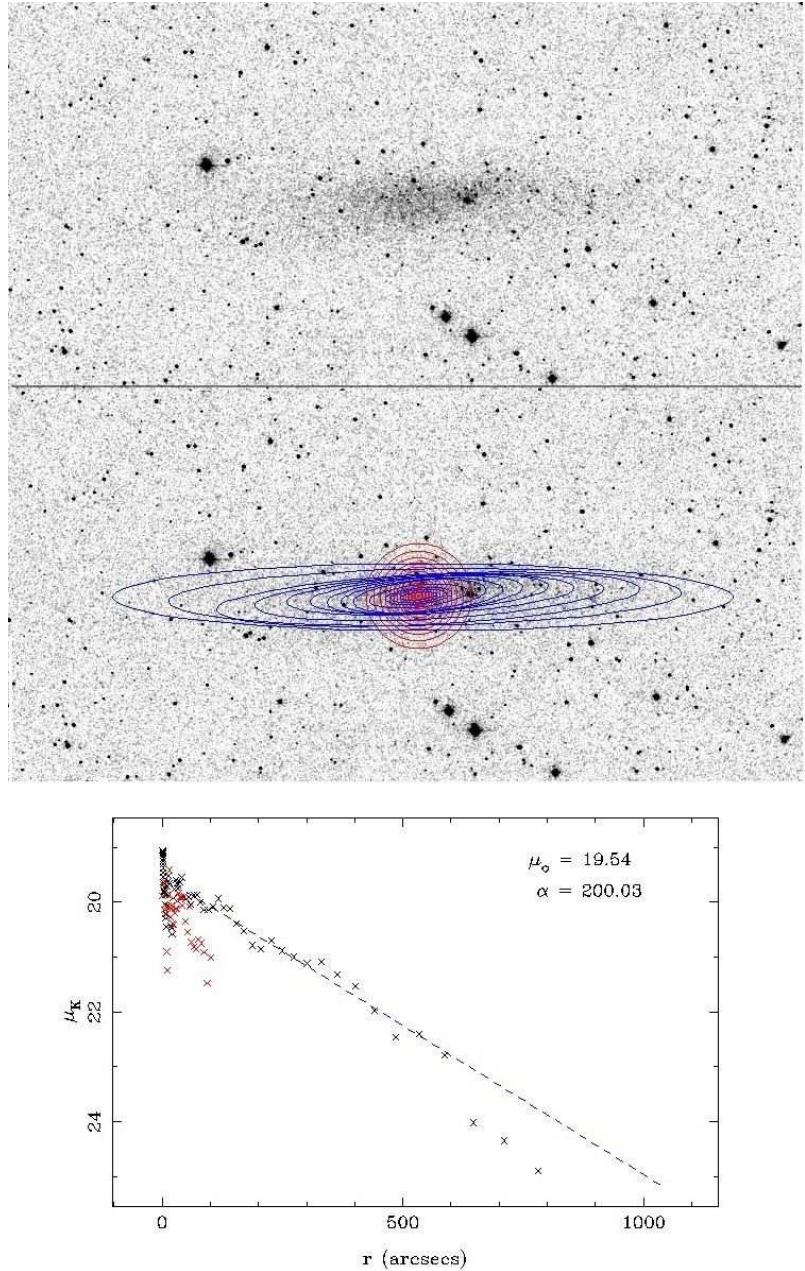}
\caption{The Sm galaxy, NGC 3109 selected from the 2MASS Large Galaxy
Atlas.  The top and middle panels display greyscale images from the 2MASS
$J$ scans.  The red ellipses are fits from the 2MASS galaxy pipeline.  The
blue ellipses are the resulting fits from this package.  The 2MASS galaxy
pipeline, for some strange reason, assumed a circular shape to was is
clearly an elongated galaxy.  This results in an underestimate of these
isophote's intensity values, as seen in the surface brightness profile in
the bottom panel.
}
\end{figure}

Given that the light is averaged around the ellipse, this effect may be
minor if the galaxy is fairly smooth and uniform.  However, galaxies that
are smooth and regular are a minority in the local Universe.  For the three
examples, shown in Figure 10, 11 and 12, the 2MASS fits consistently
underestimate the amount of disk light per isophote, as seen in comparison to
the luminosity profile determined from the raw data using the ARCHANGEL
routines.  This, in turn, results in fitted central surface brightnesses
that are too bright in central surface brightness, and fitted disk scale
lengths that are too shallow. In fact, for 49 galaxies in common between
the near-IR RSA sample (Schombert 2007) and the 2MASS Large Galaxy Atlas,
Figure 13 displays the difference between the fitted disk scale lengths
($\alpha$) and the difference between the fitted disk central surface
brightness ($\mu$).  Given the typical $\alpha$'s, the error in 2MASS fits
corresponds to a 50\% error in a galaxy's size.  Likely, errors in the
central surface brightness fits averages around 0.5 mags.  Thus, not using
the proper reduction technique not only increases the noise in the measured
parameters, but produces a biased result.

\begin{figure}
\centering
\includegraphics[width=17cm]{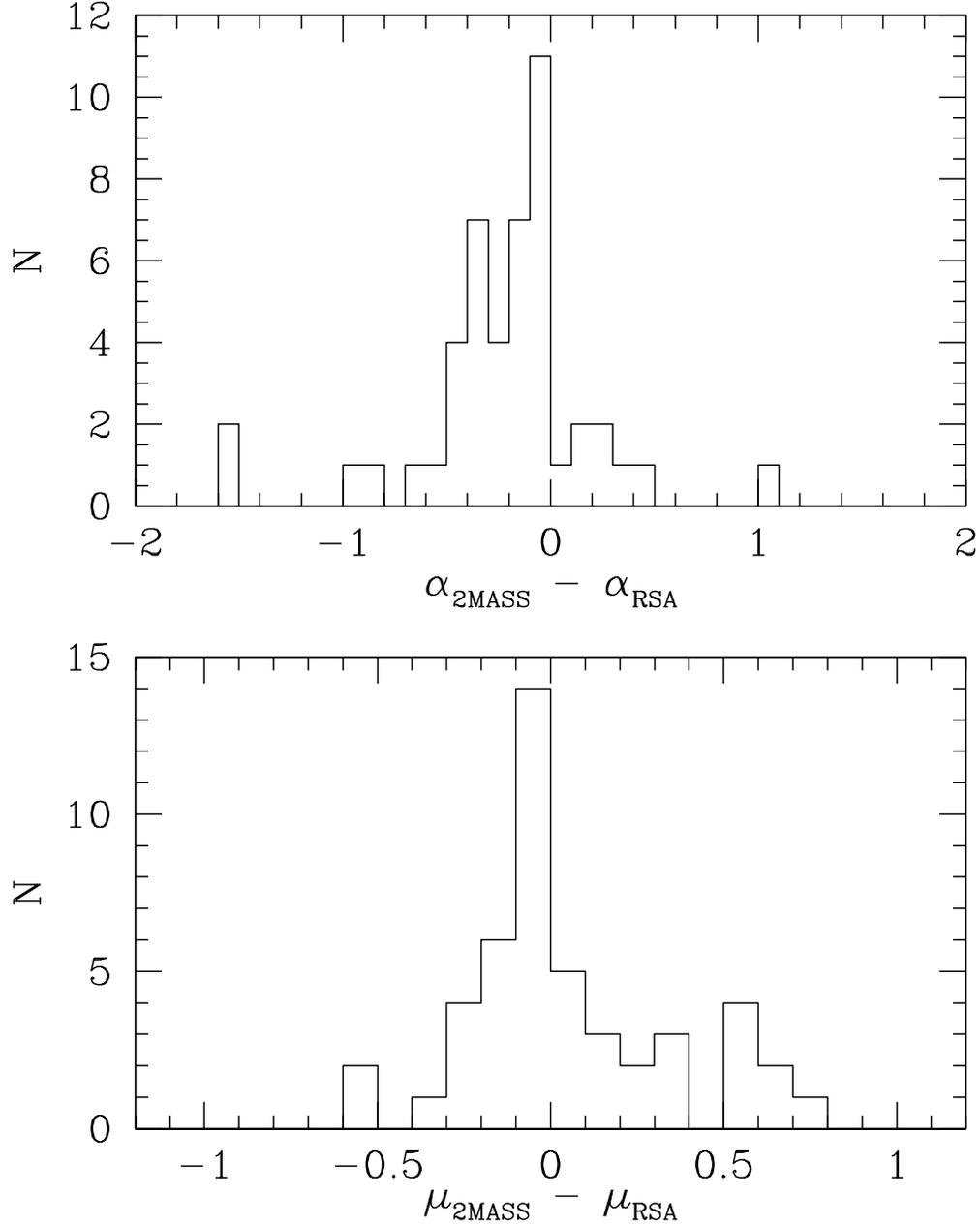}
\caption{Comparison histograms for 49 galaxies common
between the 
2MASS Large Galaxy Atlas and the new near-IR 2MASS RSA
dataset.  The top panel displays difference between disk
scale lengths ($\alpha$) from fits of an exponential law
to 2MASS surface photometry versus this package's
surface brightness reduction.  The bottom panel displays the
difference in fitted center surface brightnesses between
the 2MASS profiles and the new RSA sample.  Typically,
the 2MASS pipeline overestimates the eccentricity (too
round) which results in smaller scale lengths and
brighter central surface brightnesses.
}
\end{figure}

\section{Network Tools}

One of the more powerful modules to the Python language is the $urllib$, the
module that allows Python scripts to download any URL address.  If address is a
web page, there also exist several addition modules that parse HTML and convert
HTML tables into arrays.  This means a simple script can be written to pull down
a web page, parse it HTML and extract a data into table format.  And, on top of
this procedure, the information could be then be used into a standard GET/POST
web form used by many data archives.

As an example, the package contains {\it dss\_read}, a script that takes the
standard name for a galaxy, queries NED for its coordinates and then goes to the
DSS website and extracts the PSS-II image of the galaxy.  While this sounds like
a computationally intense task, in fact the script is composed of 49 lines.  The
downside to this network power is, of course, the possibility of abuse.
Unrestricted application of such scripts will overload websites and given network
speeds, the typical user doesn't need their own personal digital sky at their
home installation.

Lastly, various archives, in order to slow massive downloads, have an
ID/password interface.  To penetrate these sites requires the {\it mechanize}
module which simulates the actions of a brower, following links, parsing ID's and
passwords and handling cookies.  While these avatars are simple to build, the
wise usage of them remains a key challenge for the future.

\section{Package Summary}

The fastest way to learn a data reduction process is to jump in and try it.
To this end, the tarball contains all the images discussed in this
document, and several test images with known output.  This allows the user
to practice on images where the final results are known.  Thus, we
encourage the readers to download, compile and run!  Tarballs are found at
http://abyss.uoregon.edu/$\sim$js/archangel.

Another option, for the user who doesn't wish to set-up the package on
their own system (or perhaps only has a handful of galaxies to reduce), is
the client/server version of this package available at
http://http://abyss.uoregon.edu/$\sim$js/nexus (see Figure 14).  Although more
limited in its options, the web version has the advantage of speed (it's
run on a Solaris Sun Blade) and a fast learning curve.

\begin{figure}
\centering
\includegraphics[scale=0.40]{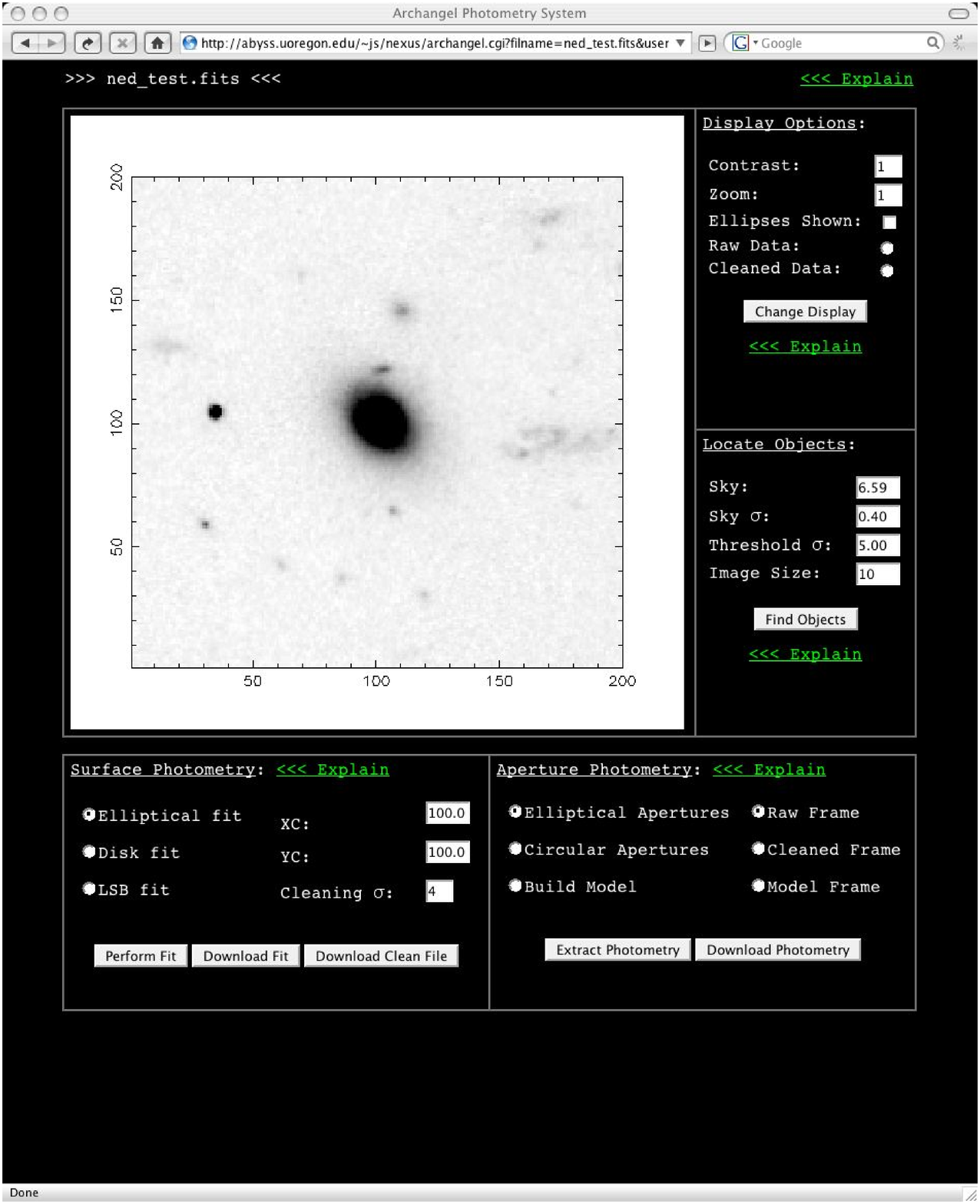}
\caption{A snapshot of the client/server version of this
package (using Firefox on a Mac OS X system).  The
ability to fit ellipses and perform aperture photometry
exists, as well a crude graphic interface to explore the
isophotal quality.
}
\end{figure}

As to the future, a number of tools need to be added to this package.  For
example, quantitative morphology uses the concentration and asymmetry
indices to parameterize a galaxy's global structure.  While these values
are easy to extract from small angular size objects, they are a challenge
for large systems.  Yet, a detailed comparison of these values to visual
morphology is a key step in understanding quantitative morphology at higher
redshifts.  However, in order get the current tools out to the
community, the package is frozen.  Additional tools
will be added to the package website as, in order of priority, 1) needed by
the PI to meet various science goals, 2) requested by outside users to
obtain their science goals, and 3) requested by outside users as possible
new computational areas to explore.  As with all evolving software, an
interested user should contact the author to see where future directions
lie (js@abyss.uoregon.edu).

\acknowledgements

This project was funded by Joe Bredekamp's incredible NASA's AIRS Program.  I am
grateful to all the suggestions I have gotten from AIRS PI's at various workshops
and panel reviews.  The program is a mixed of technology plus science types and is one
of NASA's true gems for innovative research ideas.

\appendix

\section{Package Management}

This package is a combination of FORTRAN and Python routines.  The choice of
these languages was not arbitrary.  Python is well suited for high level command
processing and decision making.  It is a clear and expressive language for text
processing.  Therefore, its style is well suited to handling file names
and data structures.  Since it is a scripting language, it is extremely portable 
between OS's.  Currently, every flavor of Unix (Linux, Mac OS X and Solaris) comes packaged
with Python.  In addition, there is a hook between the traditional astronomy
plotting package (PGPLOT) and Python (called ppgplot), which allows for easy GUI
interfaces that do not need to be compiled.

The use of FORTRAN is driven by the fact that many of the original routines for
this package were written in FORTRAN.  For processing large arrays of numbers,
C++ provides a faster routine, but current processor speeds are such that even a
2048x2048 image can be analyzed with a FORTRAN program on a dual processor
architecture faster than the user can type the next command.  STScI provides
a hook to FITS formats and arrays (called pyfits and numarray), but Python
is a factor of 100 slower than FORTRAN for array processing.

Currently there are three FORTRAN compilers in the wild, g77, gfortran and g95.
The routines in this package can use any of these compilers plus a version of
Python greater than 2.3.  CFITSIO is required and avaliable for all OS's from
its GSFC website.  The Python libaries pyfits and numarray are found at STScI's
PyRAF website.  For any graphics routines, the user will need a verson of PGPLOT
and install ppgplot as a Python library.  The ppgplot source is avaliable at the
same website as this package.  The graphics routines are only needed for data
inspection, the user should probably develop their own high-level graphics to
match their specifics.  In the directory /util one can find all the Python
subroutines to fit 1D data surface or aperture photometry.  The examples in this
manual will guide you in constructing your own interface.

Lastly, the output data files for this package are all set in XML format.  This
format is extremely cumbersome and difficult to read (it is basically an extension
of the HTML format that web browsers use).  However, a simple command line
routine is offered ($xml_archangel$) that will dump or add any parameter or
array out of or into a XML file.

\section{Core Analysis Routines}

To go from a raw data frame containing a galaxy image to a final stage
containing ellipse fits, surface photometry, profile fits and aperture values
requires three simple scripts, {\it profile}, {\it bdd} and {\it el}.  
The scripts {\it profile} and {\it el} are automatic and can be run as batch
jobs.  {\it bdd} is an interactive routine to fit the surface photometry and
is a good mid-point to study the results of the ellipse fitting.  In a
majority of cases, the user simply needs to run those three scripts with
default options to achieve their science results.

Note that command -h will provide a short summary of the commands usage.

\noindent {\it sky\_box}: \begin{verbatim}
Usage: sky_box option file_name box_size prf_file
  
options: -h = this message
         -f = first guess of border
         -r = full search, needs box_size
              and prf_file
         -t = full search, needs box_size
         -c = find sky for inner region (flats)
              needs x1,x2,y1,y2 boundarys
  
Output: 1st mean, 1st sig, it mean, it sig npts, iterations
\end{verbatim}

\noindent {$efit$}:\begin{verbatim}
Usage: efit option file_name output_file other_ops

Ellipse fitting routine, needs a standard FITS file,
output in .prf file format (xml_archange converts this
format into XML)
  
options: -h = this mesage
         -v= output each iteration
         -q = quiet
         -xy = use new xc and yc
         -rx = max radius for fit
         -sg = deletion sigma (0=no dets)
         -ms = min slope (-0.5)
         -rs = stopping radius
         -st = starting radius
 
when deleting, output FITS file called file_name.jedsub\end{verbatim}

\noindent {$prf\_smooth$}:
\begin{verbatim}
Usage: prf_smooth option prf_file_name

parameter #6 set to -1 for cleaned ellipse, 0 for
unfixable ones

options: -x = delete unfixable ones
         -s = spiral, low smooth
         -d = neutral smooth
         -q = quick smooth
\end{verbatim}

\noindent {$prf\_edit$}: \begin{verbatim}
Usage: prf_edit file_name

visual editor of isophote ellipses output from efit

note: needs a .xml file, works with cleaned images

cursor commands:
r = reset display      z = zoom in
c = change cont        x = flag ellipse/lum point
t = toggle wd cursor   o = clean profile
q = exit               h = this message
\end{verbatim}

\noindent {$probe$}: \begin{verbatim}
Usage: probe option master_file

quick grayscale display GUI

options: -f = do this image only
         -m = do file of images

cursor commands:
/ = abort              q = move to next frame
c = contrast           r = reset zoom
z = zoom               t = toggle ellipse plot
p = peek at values
a,1-9 = delete circle  b = delete box
\end{verbatim}

\noindent {$bdd$}: \begin{verbatim}
Usage: bdd options file_name

quick surface photometry calibration and fitting GUI

options: -h = this message
         -p = force sfb rebuild

window #1 cursor commands:
c = contrast control   r = reset boundaries
z = zoom on points     x = delete point
s = set sky (2 hits)   i = show that ellipse
/ = write .sfb file    q = abort

window #2 cursor commands:
x = erase point        d = disk fit only
m = erase all min pts  f = do bulge+disk fit
u = erase all max pts  e = do r**1/4 fit only
b = redo boundaries    p = toggle 3fit/4fit
q = abort              r = reset graphics
/ = write .xml file and exit

\end{verbatim}

\noindent {\it xml\_archangel}: \begin{verbatim}
   xml_archangel op file_name element data

   add or delete data into xml format
      -o = output element value or array
      -d = delete element or array
      -a = replace or add array, array header and data is cat'ed into routine
      -e = replace or add element
      -c = create xml file with root element
      -k = list elements, attributes, children (no data)
\end{verbatim}

\noindent {$el$}: \begin{verbatim}
Usage: el options cleaned_file

script that takes cleaned FITS file and fills in NaN
pixels
from efit isophotes, then does elliptical apertures on
resulting .fake file

options: -v = verbose
\end{verbatim}

\noindent {$asymptotic$}: \begin{verbatim}
Usage: asymptotic xml_file

simply GUI that determines asymptotic fit on integrated
galaxy mag,
delivers mag/errors from apertures and curve of growth
fit into XML file

cursor commands:
r = reset       a = adjust lum for better fit
f = linear fit  x,1,2,3,4 = delete points
z = set profile extrapolation point
/ = exit        b = change borders
\end{verbatim}

\end{document}